# Estimation of the excess mortality in chronic diseases from prevalence and incidence data

Ralph Brinks

## Introduction

Aggregated data such as health insurance claims data become more and more available for research purposes. Recently, we proposed a new method to estimate the excess mortality in chronic diseases from aggregated age-specific prevalence and incidence data [Toe18, Bri19]. So far, estimates of excess mortality have only been possible for ages 50+ and have shown to be unstable in younger ages. For example, in the simulation study of [Bri19], the bias increases as the age decreases (Table 1 in [Bri19]).

The aim of this article is to explore the reasons why estimates of excess mortality for younger ages are prone to bias and what can be done to extend the age range to ages below 50 years. As a testing example, we use claims data about diabetes from the German statutory health insurance based on about 70 million people collected during the period from 2009 to 2015 [Gof17].

## Methods

It can be shown that the temporal change, $\partial p = (\partial_t + \partial_a)\, p$ of the age-specific prevalence $p$ is related to the incidence rate $i$, the mortality rates $m_0$ and $m_1$ of the people with and without the disease, respectively, the general mortality $m$ and the mortality rate ratio $R = m_1/m_0$ via the following equations [Bri14, Bri16]:

$$\partial p = (1 - p)\,\{i - p \times (m_1 - m_0)\} \qquad (1a)$$

$$= (1 - p)\,\{i - m \times p\,(R - 1) / [1 + p\,(R - 1)]\}. \qquad (1b)$$

Given the age-specific prevalence $p$, the age-specific incidence rate $i$ and the general mortality rate $m$, Equations (1a) and (1b) can be used to estimate the excess mortality rate $\Delta m = m_1 - m_0$ and the mortality rate ratio $R$:

$$\Delta m = \{i - \partial p/(1 - p)\} / p, \tag{2a}$$

$$R = 1 + 1/p \times \{i - \partial p/(1 - p)\} / \{m - i + \partial p/(1 - p)\}. \tag{2b}$$

Goffrier and colleagues report the age-specific prevalence $p$ of type 2 diabetes for men in 2009 and 2015 [Gof17]. Furthermore, the age- and sex-specific incidence rate $i$ in middle of the period, i.e., in the year 2012, is reported. These data are used as input for Equations (2a) and (2b). In addition, for applying Equation (2b) we use the general mortality $m$ in 2012 from the Federal Statistical Office of Germany [Fed19].

To obtain estimates for the excess mortality $\Delta m$ and the mortality rate ratio $R$, we essentially follow two directions: First, we apply Equations (2a) and (2b) directly to the input data (direct approach). In a second – indirect – approach, we choose candidate values for the mortality rate ratio $R$, solve Equation (1b) and compare the computed solution for the candidate $R$ with the observed solution in terms of the squared differences. Both methods, direct and indirect are detailed in the next paragraphs.

## *Direct methods*

The age-specific prevalence $p$ for men in 2009 and 2015 are modeled by a linear regression model after application of a logit transformation. The age-specific incidence rate $i$ for 2012 is modeled by a linear regression model after a log-transformation. In the direct estimation approach, we then seek to apply Equations (2a) and (2b). However, possible sampling uncertainties in the input data may lead to uncertainties in the estimated excess mortality rates $\Delta m$ and the mortality rate ratio $R$. To take these sampling uncertainties into account, the input data (prevalence $p$ in 2009 and 2015, incidence $i$ in 2012) have been randomly resampled 2000 times with respect to a binomial error model for $p$ and a Poisson error model for $i$. For each resampled combination, Equations (2a) and (2b) are applied, each yielding a replicate result for $\Delta m$ and $R$. Based on these 2000 replicate results, the median, the 2.5 and the 97.5 percentiles of age-specific $\Delta m$ and $R$ are calculated and reported.

*Indirect method*

For the indirect method, we choose candidate values for the logarithm of the mortality rate ratio $R$ at ages 30, 60, and 90 years. For brevity, let $x$ denote the three dimensional vector $x = (\log(R(30), \log(R(60)), \log(R(90)))$. Ages in the intervals (30, 60) and (60, 90) are interpolated by a straight line on the log-scale. The rationale for this choice of $R$ is based on the idea that $m_0$ and $m_1$ follow a Gompertz-Makeham law [Mis13]. Then, the logarithm of the quotient $m_1/m_0$ is a straight line. Ages below 30 and above 90 are extrapolated by $R(30)$ and $R(90)$, respectively. Then, we solve the PDE (1b) using $i$ and $m$ with initial condition $p(2009; a)$. The incidence rate $i$ and the initial condition $p(2009, a)$ are given by [Gof17]. The general mortality stems from the Federal Statistical Office of Germany [Fed19]. By solving the PDE (1b), we obtain a solution $p_c$ of the prevalence in 2015 given the candidate values of the $R$ at ages 30, 60, and 90. This solution $p_c$ in 2015 is compared with the surveyed values $p_{obs}(2015, a)$ in terms of the summed squared differences $(p_{obs} - p_c)^2$.

For numerically solving the PDE (1b), the PDE is first converted into an ordinary differential equation (ODE) by the method of characteristics [Pol] and then, the associated ODE is solved by the Runge-Kutta Method of fourth order [Dah74].

The rationale behind the indirect approach is the observation that direct application of (2a) and (2b) may lead to numerically instable results. Indications for a possible numerical instability may be seen in the ill-posedness of the inverse problem [Bri18].

# Results

*Direct methods*

Figure 1 shows the median age-specific excess mortality $\Delta m$ (solid line) and the associated 2.5 and 97.5 percentiles (dashed lines) based on the 2000 replicate results from application of Equation (2a). At ages below 40 years, the difference between the 2.5 and 97.5 percentiles is greater than in ages above 40 years. One may see a funnel widening up towards lower ages. Thus, the amount of uncertainty increases as the age decreases.

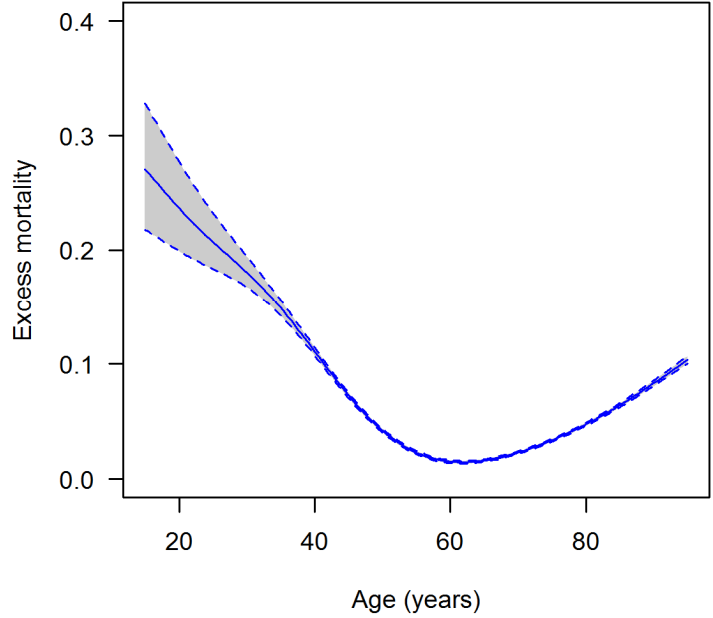

**Figure 1: Age-specific excess mortality. Median (solid line), 2.5 and 97.5 percentiles (lower and upper dashed lines) based on 2000 replicate results.**

Application of Equation (2b) for obtaining the mortality rate ratio $R$, yields the results as shown in Table 1. We see that for ages below 55 years of age, the mortality rate ratios are implausibly high or even turn negative. By definition of the mortality rate ratio, a quotient of two positive rates, negative values are not possible. Thus, we see that direct estimates based on Equation (2b) do not yield sensible results for lower age groups and are not reliable.

**Table 1: Mortality rate ratios ($R$) for different age-groups. The age group is indicated in the top row, the rows below are the 2.5 percentile, the 50 percentile (median), and the 97.5 percentile based on 2000 replicate results.**

|      | 15-19 | 20-24 | 25-29  | 30-34 | 35-39 | 40-44 | 45-49 | 50-54 | 55-59 | 60-64 | 65-69 | 70-74 | 75-79 | 80-84 | 85-89 | 90+ |
|------|-------|-------|--------|-------|-------|-------|-------|-------|-------|-------|-------|-------|-------|-------|-------|-----|
| 2.5  | 644   | 550   | -11424 | -347  | -130  | -90   | -6987 | 12.5  | 4.0   | 2.7   | 2.5   | 2.3   | 2.1   | 1.9   | 1.6   | 1.5 |
| 50   | 850   | 745   | -3409  | -330  | -126  | -83   | -444  | 14.6  | 4.4   | 2.9   | 2.6   | 2.4   | 2.1   | 1.9   | 1.7   | 1.5 |
| 97.5 | 1111  | 1023  | -2028  | -315  | -122  | -78   | 7037  | 17.2  | 4.9   | 3.1   | 2.7   | 2.5   | 2.2   | 1.9   | 1.7   | 1.5 |

## *Indirect method*

Figure 2 shows the Euclidean norm of $x = (\log(R(30), \log(R(60)), \log(R(90))))$ over the summed squared differences $(p_{obs} - p_c(x))^2$. Each blue dot corresponds to one pair of $x$ and $(p_{obs} - p_c(x))^2$. In the lower left part of the graph (towards the origin), there is a region without any blue dot. The boundary between this white and blue region forms a curve that is shaped like an L and correspondingly is called L-curve. The L-curve is well-known in ill-posed inverse problems [Han99, Han07].

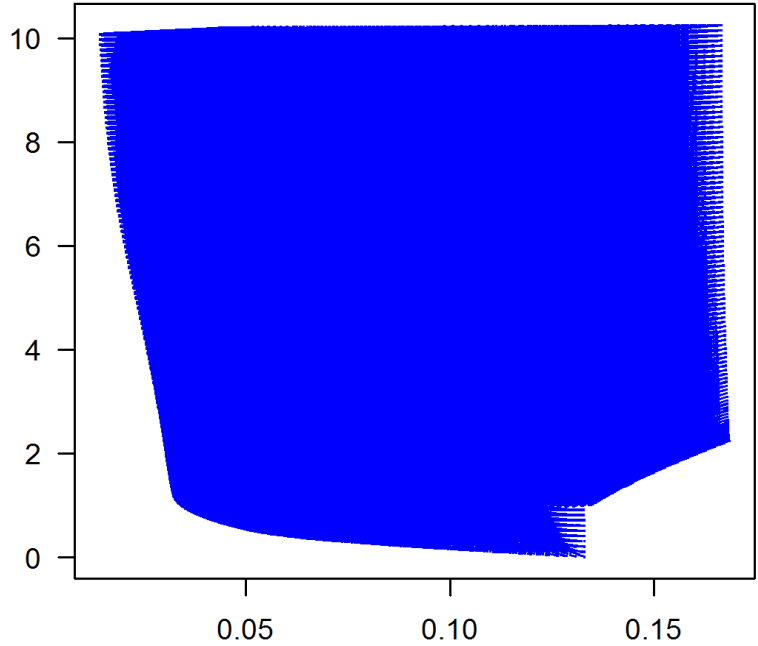

Figure 2: Norm of $x = (\log(R(30)), \log(R(60)), \log(R(90)))$ (ordinate) over the summed squared differences $(p_{obs} - p_c(x))^2$.(abscissa). Each blue dot corresponds to one pair of $x$ and $(p_{obs} - p_c(x))^2$. The boundary of the blue region to left and to the bottom forms an L-shaped curve called the L-curve [].

## Conclusion

In this manuscript we have described methods to estimate two epidemiological indices for the excess mortality of a chronic condition from age-specific prevalence and incidence data. The first index is the rate difference $\Delta m$ between the mortality rate of the diseased people ($m_1$) and the people without the disease ($m_0$), i.e., $\Delta m = m_1 - m_0$. The second index is the mortality rate ratio $R = m_1/m_0$. In an example about diabetes in Germany, it turns out that estimates based on the mortality rate ratio $R$ are numerically unstable and can lead to implausible results. A possible reason lies in the ill-posedness of the underlying inverse problem. An indication for this reason is the observed L-curve resulting from the indirect method, which is typical for ill-posed inverse problems.

From the current findings, we would recommend to estimate figures of the excess mortality on the mortality rate difference $\Delta m$ instead of the mortality rate ratio $R$. In the example application, estimates for the rate difference $\Delta m$ yield sensible results.

Contact
Ralph Brinks
German Diabetes Center
Institute for Biometry and Epidemiology
University Duesseldorf
40225 Duesseldorf
Germany